\documentclass[12pt]{JHEP3}
\usepackage{epsfig}
\epsfclipon
\usepackage{multicol}
\usepackage{epsfig,bm}
\usepackage{amssymb,amsmath}
\usepackage{graphicx}
\newcommand{\be}{\begin{equation}}
\newcommand{\ee}{\end{equation}}
\newcommand{\bea}{\begin{eqnarray}}
\newcommand{\beas}{\begin{eqnarray*}}
\newcommand{\eea}{\end{eqnarray}}
\newcommand{\eeas}{\end{eqnarray*}}
\newcommand{\ba}{\begin{array}}
\newcommand{\ea}{\end{array}}
\def\ls{\mathrel{\lower4pt\vbox{\lineskip=0pt\baselineskip=0pt
           \hbox{$<$}\hbox{$\sim$}}}}
\def\gs{\mathrel{\lower4pt\vbox{\lineskip=0pt\baselineskip=0pt
           \hbox{$>$}\hbox{$\sim$}}}}
\def\smiley{\hbox{\large$\bigcirc$\hspace{-.80em}%
\raise.2ex\hbox{$\cdot\cdot$}\kern-.61em    %--- .56
\lower.2ex\hbox{\scriptsize$\smile$}}\ }

\newcommand{\roughly}[1]{\mathrel{\raise.3ex\hbox{$#1$\kern-0.85em
\lower1ex\hbox{$\sim$}}}}

\def\be{\begin{equation}}
\def\beq\begin{equation}
\def\ee{\end{equation}}
\def\bea{\begin{eqnarray}}
\def\eea{\end{eqnarray}}

\def\beq{\begin{equation}}
\def\eeq{\end{equation}}
\def\beqa{\begin{eqnarray}}
\def\eeqa{\end{eqnarray}}

\newcommand{\bmat}{\left(\begin{array}}
\newcommand{\emat}{\end{array}\right)}

\title{A-term inflation and the smallness of the neutrino masses}

\author{Rouzbeh Allahverdi$^{1,2}$,
Alexander Kusenko$^{3}$,
Anupam Mazumdar$^{4}$,\\
$^{1}$~Perimeter Institute for Theoretical Physics, Waterloo, ON,
N2L 2Y5, Canada \\
$^{2}$~Department of Physics and Astronomy, McMaster University, Hamilton,
ON, L8S 4M1, Canada\\
$^{3}$~Department of Physics and Astronomy, University of California, Los
Angeles, CA 90095-1547\\
$^{4}$~NORDITA, Blegdamsvej-17, Copenhagen-2100, Denmark}

\abstract{
The smallness of the neutrino masses may be related to inflation.  The
minimal supersymmetric Standard Model (MSSM) with small neutrino Yukawa
couplings already has all the necessary ingredients for a successful
inflation. In this model the inflaton is a gauge-invariant combination
of the right-handed sneutrino, the slepton, and the Higgs field, which
generate a flat direction suitable for inflation if the Yukawa
coupling is small enough. In a class of models, the observed microwave
background anisotropy and the tilted power spectrum are related to the
neutrino masses.}

\begin{document}

%%%%%%%%%%%%%%%%%%%%%%%%%%%%%%%%%%%%%%%%%%%%%%%%%%%%%%%%%%%%%%%%%
\noindent

One of the least satisfying features of inflation is the need to
introduce the inflaton as an {\em ad hoc} ingredient.  However, some
well-motivated theories beyond the Standard Model may already have the
gauge-invariant scalar degrees of freedom necessary for inflation.
One recent example is the use of a flat direction in the minimal
supersymmetric Standard Model (MSSM) as an inflaton~\cite{AEGM}. The
model illustrates a successful inflation at low scales whose
ingredients are testable at Large hadronic Collider (LHC)~\cite{LHC}
and CMB~\cite{WMAP3}.  There are a few other examples of using some
gauge-invariant degrees of freedom already present in the theory for
the purposes of inflation\cite{FEW}.

In any model, the inflaton potential must be very flat, which is
suggestive of either a symmetry or a small coupling, or both.  We will
see that this property of the inflaton may be related to the smallness
of neutrino masses.  Identifying such a connection could have
important ramifications and could lead to a more fundamental theory.
The model we will use as an example will contain nothing but the MSSM
and the right-handed neutrinos.  We will show that a viable inflation
in this model favors the correct scale for the neutrino masses.

Let us consider the minimal supersymmetric standard model (MSSM) with
three additional fields, namely the right-handed (RH) neutrino
supermultiplets. The relevant part of the superpotential
is

\beq \label{supot}
W = W_{\rm MSSM} + {\bf h} {\bf N} {\bf H}_u {\bf L}.
\eeq
Here ${\bf N}$, ${\bf L}$ and ${\bf H}_u$ are superfields containing
the RH neutrinos, left-handed (LH) leptons and the Higgs which gives
mass to the up-type quarks, respectively. For conciseness we have
omitted the generation indices.  We note that the RH (s)neutrinos are
singlets under the standard model (SM) gauge group. However in many
extensions of the SM they can transform non-trivially under the action
of a larger gauge group.  The simplest example is extending the SM
gauge group to $SU(3)_{\rm C} \times SU(2)_{\rm W} \times U(1)_{\rm Y}
\times U(1)_{\rm B-L}$, which is a subgroup of $SO(10)$.  Here $B$ and
$L$ denote the baryon and lepton numbers, respectively. This is the
model we consider here.  In particular, the $U(1)_{\rm B-L}$ prohibits
the RH Majorana masses.

There is a broad range of phenomenologically acceptable scales
$\Lambda_{\rm B-L}$ at which the $U(1)_{\rm B-L}$ can be broken.  In
the low-energy effective theory this breaking can introduce the RH
Majorana mass for neutrinos $W \supset {M \over 2} {\bf N} {\bf N}$,
where $M \propto {\Lambda}_{\rm B-L}$.

The active neutrino masses that arise from this are given by the usual
seesaw relation $h^2 {\langle H_u
\rangle}^2/M$~\cite{seesaw,NEUT-REV}, where $\langle H_u \rangle$ is
the Higgs vacuum expectation value (VEV). Although the seesaw
mechanism {\em allows} for small neutrino masses in the presence of
large Yukawa couplings, it {\em does not require } the Yukawa
couplings to be of order one.  It still allows one to choose between
the large Yukawa couplings and large Majorana masses on the one hand,
and the small Yukawas, small Majorana masses, on the other hand.
Viable models for neutrino mass matrices have been constructed in both
limits, including the low-scale seesaw models~\cite{deGouvea:2005er,
nuMSM}, in which the Yukawa couplings are typically of the order of
\beq \label{yukawa}
h \sim 10^{-12},
\eeq
or the same order of magnitude, as it would have in the case of Dirac
neutrinos in order to explain the mass scale $\sim {\cal O}(0.1~{\rm
eV})$ corresponding to the atmospheric neutrino oscillations detected
by Super-Kamiokande experiment. As shown in Ref.~\cite{nuMSM}, such
models can explain all the observed data on the neutrino masses and
mixing, and, in addition, one can have dark matter in the form of
sterile neutrinos with mass of the order of several
keV~\cite{dark_matter}. Moreover, in some range of parameters, one can
also explain the observed velocities of pulsars\cite{pulsars} and the
baryon asymmetry of the universe~\cite{baryogenesis}.  In addition,
the x-ray background produced by decays of the relic sterile neutrinos
can play a role in the formation of the first stars that lead to the
reionization~\cite{reion}.

In view of the viability of the small-Yukawa scenario for the neutrino
masses, let us consider the flat directions associated with such small
Yukawa couplings in connection with cosmological inflation.

The scalar potential of the MSSM has numerous flat directions made up
of squarks, sleptons, and the Higgs fields~\cite{DRT} (for a review
see~\cite{MSSM-REV}). These directions are classified by monomials
which are invariant under the SM gauge group. Those monomials with
$B-L = 0$ will be also $D$-flat under $U(1)_{\rm B-L}$, while those
with $B-L \neq 0$ must be multiplied by an appropriate number of ${\bf
N}$ superfields. In particular, ${\bf N} {\bf H}_u {\bf L}$ is now a
$D$-flat direction. We note that, in the absence of a gauged
$U(1)_{\rm B-L}$, there would be two independent flat directions:
${\bf N}$ and ${\bf H}_u {\bf L}$. These flat directions could develop
different VEVs, could start oscillating at different times, and would
decay independently from each other.  However, when the SM gauge
symmetry is extended to include a $U(1)_{\rm B-L}$, the true $D$-flat
direction is ${\bf N} {\bf H}_u {\bf L}$. As we shall see, this is
crucial for the proposed model of inflation and a successful
reheating. Such a gauged $U(1)_{\rm B-L}$ which is broken just above
TeV is compatible with all phenomenological constraints.

Let us now work in the basis where neutrino masses are
diagonalized. There is a flat direction ${\bf N}_3 {\bf H}_u {\bf
L}_3$ spanned by the VEV of the lower and upper weak isospin
components of ${\bf H}_u$ and ${\bf L}_3$, respectively.
The scalar field corresponding to the flat direction is denoted by
\beq \label{flat}
\phi = {{\tilde N}_3 + H^2_u + {\tilde L}^1_3 \over \sqrt{3}}\,,
\eeq
where the superscripts refer to the weak isospin components.  One must
now include the soft SUSY breaking terms, such as the mass terms and
the $A$-term. The $A$-terms are known to play an important role in
Affleck-Dine baryogenesis~\cite{DRT}, as well as in the inflation
models based on supersymmetry~\cite{curvaton,AEGM,DHL}.

The potential along the flat direction is found to be
\begin{eqnarray} \label{flatpot}
V (\phi) = \frac{m^2_{\phi}}{2} \phi^2 +
\frac{h^2_3}{12} \phi^4 \,
+ \frac{A h_3}{6\sqrt{3}} {\rm cos}
\left(\theta + \theta_h + \theta_A \right) \phi^3 \,,
\end{eqnarray}
where the flat direction mass is given in terms of the soft masses of
${\tilde N_3},~H_u$, and ${\tilde L_3}$: $m^2_{\phi} =
\left(m^2_{\tilde N_3} + m^2_{H_u} + m^2_{\tilde L_3}\right)/3$. Here
we have used the radial and angular components of the flat direction
$\phi_R + i \phi_I = \sqrt{2}\phi \, {\rm exp} \left(i \theta \right)$
, and $\theta_h,~\theta_A$ are the phases of the Yukawa coupling $h_3$
and the $A$-term, respectively. We note that the above potential does
not contain any non-renormalizable term at all.

\begin{figure}
\vspace*{-0.0cm}
\begin{center}
\epsfig{figure=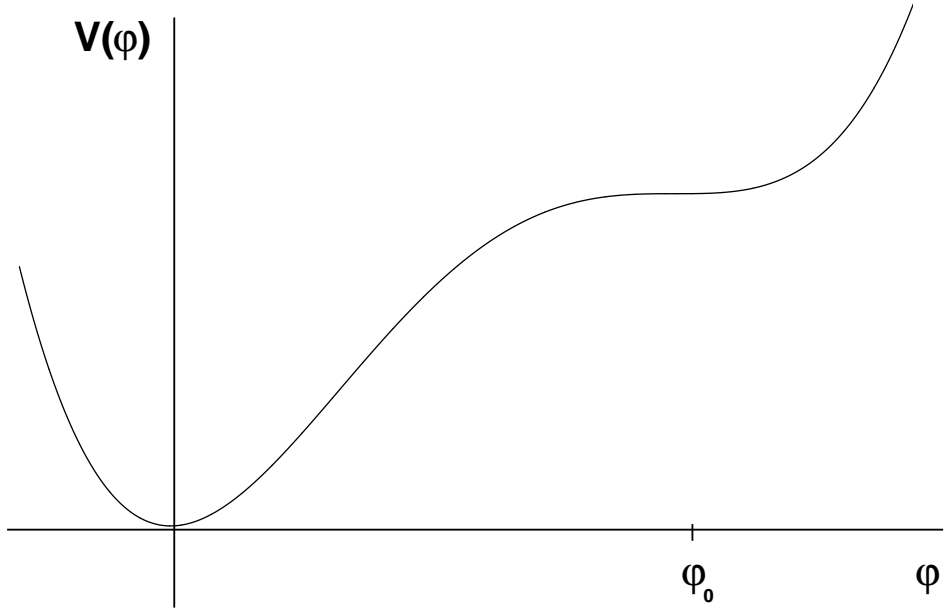,width=.84\textwidth,clip=}
\vspace*{-0.0cm}
\end{center}
\caption{ The inflaton potential. The potential is flat near the
saddle point where inflation occurs.}
\label{plot}
\end{figure}

The last term on the right-hand side of eq.~(\ref{flatpot}) is
minimized when ${\rm cos} \left(\theta + \theta_h + \theta_A \right)
= -1$. Along this direction, $V(\phi)$ has the global minimum at
$\phi=0$ and a local minimum at $\phi_0 \sim m_{\phi}/h_3$, as long
as
\beq \label{extrem} 4 m_\phi \leq A \leq 3 \sqrt{2} m_\phi \,.
\eeq
If $A > 3 \sqrt{2} m_{\phi}$, the minimum at $\phi_0$ will become
global. However, this will not be relevant for our discussion below.
As we will point out later, successful inflation is only possible if
$A$ is very close to $4 m_{\phi}$, which ensures $\phi_0$ being a
local minimum. At the local minimum the curvature of the potential
along the radial direction is $\sim + m^2_{\phi}$, and the curvature
along the angular direction is positive.  Near $\phi=\phi_0$ the
potential reduces to: $V \sim {m^4_{\phi}/ h^2_3}$.  If $\phi$ is
trapped in the false vacuum and its potential energy $V$ dominates
the total energy density of the Universe, then inflation ensues. The
Hubble expansion rate during inflation is given by
\beq \label{hubble}
H_{inf} \sim {m_{\phi} \phi_0 \over M_{\rm P}} \sim
{m^2_{\phi}\over h_3 M_{\rm P}} \,.
\eeq
We note that $H_{inf} \ll m_{\phi}$ as $\phi_0 \ll M_{\rm P}$. This
implies that, around the local minimum, the potential barrier is too
high for the flat direction, ${\phi}$, to jump over.  The situation is
essentially the same as that in the old inflation
scenario~\cite{GUTH}, which could yield expansion by many e-foldings,
but which suffered from the {\em graceful exit} problem.

However, the barrier disappears when the inequality in
eq.~(\ref{extrem}) is saturated, i.e., when
\beq \label{cond}
A = 4 m_{\phi} \,.
\eeq
Then both first and second derivatives of $V$ vanish at $\phi_0$,
$V^{\prime}(\phi_0)=V^{\prime\prime}(\phi_0)=0$, and the potential
becomes extremely flat in the {\it radial direction}, see
Fig.~[\ref{plot}].  We note that individually none of the terms in
eq.~(\ref{flatpot}) could have driven a successful inflation at VEVs
lower than $M_{\rm P}=2.4\times 10^{18}$~GeV. However the combined
effect of all the terms leads to a successful inflation without the
graceful exit problem~\footnote{In what follows we study inflation
when eq.~(\ref{cond}) is satisfied. However, this condition is not
strictly required. Successful inflation is obtained for values of $A/4
m_{\phi}$ which are slightly larger or smaller than $1$, $(\vert A^2 -
16 m^2_{\phi} \vert^{1/2}/4 m_{\phi}) \ls 10^{-8}$, for detailed
discussion see~\cite{AEGJM}.}.

Around $\phi_0$ the field is stuck in a plateau with potential energy
\begin{eqnarray}\label{phi0}
V(\phi_0) = {m^4_{\phi} \over 4 h^2_3} \,,~~
\phi_0 = \sqrt{3} {m_{\phi} \over h_3} \, .
\end{eqnarray}
The first and second derivatives of the potential vanish, while
the third derivative does not. Around $\phi=\phi_0$ one can expand the
potential as
$V(\phi) =V(\phi_0) + (1/ 3!)V'''(\phi_0)(\phi-\phi_0)^3$,
where
\begin{eqnarray}
\label{3rder}
V^{\prime \prime \prime}({\phi_0}) = {2 \over
\sqrt{3}} h_3 m_{\phi}\,.
\end{eqnarray}
Hence, in the range $[\phi_0 - \Delta \phi, \phi_0 + \Delta \phi]$,
where $\Delta \phi \sim H_{\rm inf}^2/V^{\prime\prime\prime}(\phi_0)
\sim \left({\phi}^3_0/M^2_{\rm P}\right) \gg H_{\rm inf}$, the potential is
flat along the real direction of the inflaton.  Inflation
occurs along this flat direction.

If the initial conditions are such that the flat direction starts in
the vicinity of $\phi_0$ with $\dot\phi\approx 0$, then a sufficiently
large number of e-foldings of inflation can be generated. Around the
saddle point, due to the random fluctuations of the massless field,
the quantum diffusion is stronger than the classical force, $H_{\rm
inf}/2\pi > \dot\phi/H_{\rm inf}$~\cite{LINDE}, for
\beq \label{drift}
{(\phi_0-\phi) \over \phi_0} \ls \Big({m_\phi \phi_0^2 \over
M_{\rm P}^3}\Big)^{1/2}
= \left({3 m^3_{\phi} \over
h^2_3 M^3_{\rm P}}\right)^{1/2}
\, .
\eeq
At later times, the evolution is determined by the usual slow roll.
The equation of motion for the $\phi$ field in the slow-roll
approximation is $3H\dot\phi
=-(1/2)V'''(\phi_0)(\phi-\phi_0)^2$.

A rough estimate of the number of e-foldings is then given by
\beq \label{efold}
{\cal N}_e(\phi) = \int {H d\phi \over \dot\phi}
\simeq \left({m_{\phi} \over 2 h_3 M_{\rm P}}\right)^2 {\phi_0 \over
(\phi_0-\phi)} ~ ,
\eeq
where we have assumed $V'(\phi) \sim (\phi - \phi_0)^2 V'''(\phi_0)$
(this is justified since $V'(\phi_0)$ and $V''(\phi_0)$ are both small). We
note that the initial displacement from $\phi_0$ cannot be much smaller than
$H_{\rm inf}$, due to the uncertainty from quantum fluctuations.

Inflation ends when the slow roll parameters become $\sim 1$. It
turns out that $\vert \eta \vert \sim 1$ gives the dominant
condition
\beq \label{end} {(\phi_0-\phi) \over \phi_0}
% \sim \Big({\phi_0\over 12 M_{\rm P}}\Big)^{1/2}
\sim {\sqrt{3} m^3_{\phi} \over 24 h^3_3 M^3_{\rm P}}\,. \eeq
The total number of e-foldings can be computed as~\cite{AEGM}:
\beq \label{totalefold}
{\cal N}_{e} \sim \left({\phi_0^2 \over m_\phi
M_{\rm P}} \right)^{1/2} = \left({3 m_{\phi} \over
h^2_3 M_{\rm P}}\right)^{1/2} \,,
\eeq
evaluated after the end of diffusion, see eq.~(\ref{drift}), when the slow-roll
regime is achieved.

Let us now consider adiabatic density perturbations. As in Ref.~\cite{AEGM},
one finds
\beq
\label{amp}
\delta_{H}\simeq \frac{1}{5\pi}\frac{H^2_{inf}}{\dot\phi}
\sim {h^2_3 M_{\rm P}\over 3 m_{\phi}}\,{\cal N}_{\rm COBE}^2\,.
\eeq
In the above expression we have used the slow roll approximation
$\dot\phi\simeq -V'''(\phi_0)(\phi_0- \phi)^2/3H_{\rm inf}$, and
eq.~(\ref{efold}).  The number of e-foldings, ${\cal N}_{\rm COBE}$,
required for the observationally relevant perturbations, is $\geq
60$~\cite{LEACH}. The exact number depends on the scale of inflation
and on when the Universe becomes radiation dominated (we note that full
thermalization is not necessary as it is the relativistic equation of
state which matters). In our case ${\cal N}_{\rm COBE}<60$ as we shall
see below.

The spectral tilt of the power spectrum and its running are
\begin{eqnarray}
\label{spect}
n_s &=& 1 + 2\eta - 6\epsilon \simeq 1 - {4\over {\cal N}_{\rm COBE}}, \\
{d\,n_s\over d\ln k} &=& - {4\over {\cal N}_{\rm COBE}^2} \,,
\end{eqnarray}
{\em cf.} \cite{AEGM}.  (We note that $\epsilon \ll 1$ while $\eta =
-2/{\cal N}_{\rm COBE}$.)

It is a remarkable feature of the model that for the weak-scale
supersymmetry and for the correct value of the Yukawa coupling,
namely,
\beq
\label{param}
m_{\phi} \simeq 100~{\rm GeV}-10~{\rm TeV}\,,~~h_3\sim 10^{-12}\,,
\eeq
the flat direction ${\bf N}_3 {\bf H}_u {\bf L}_3$ leads to a
successful low scale inflation near $\phi_0\sim
\left(10^{14}-10^{15}\right){\rm GeV} \ll M_{\rm P}$, with
\begin{eqnarray} \label{values}
&& V\sim 10^{32}-10^{36}~{\rm GeV}^4\,,~~~~H_{\inf}\sim
10~{\rm MeV}-1{\rm GeV}\,, \nonumber \\
&& {\cal N}_e\sim 10^{3}\,,~~~~T_{max}\sim 10^{8}-10^{9}~{\rm GeV}\,.
\end{eqnarray}
The total number of e-foldings driven by the slow roll inflation,
${\cal N}_e \sim 10^3$, is more than sufficient to produce a patch of
the Universe with no dangerous relics.  Those domains that are
initially closer to $\phi_0$ enter self-reproduction in eternal
inflation.  Since the inflaton, ${\bf N}_3 {\bf H}_u {\bf L}_3$,
couples directly to MSSM particles, after inflation the field
oscillates and decays to relativistic MSSM degrees of freedom.  The
highest temperature during reheating is $T_{max}\sim V^{1/4}$.  This
temperature determines the total number of e-foldings required for the
relevant perturbations to leave the Hubble radius during inflation; in
our case it is roughly ${\cal N}_{\rm COBE}\sim 50$.

An interesting observation is that the VEV is smaller than the Planck
scale, therefore, the potential is free of any {\it trans-Planckian}
corrections, and the {\it supergravity effects} are negligible since
$m_{\phi}\gg H_{\inf}$. Hence, one can make more certain predictions
than in the case of the high-scale inflation.

Despite the low scale of inflation, the flat direction can generate
density perturbations of the correct size for the parameters listed above.
Indeed, from eqs.~(\ref{amp},\ref{spect}), and (\ref{param}), we obtain:
\beq
\delta_{H}\sim 10^{-5}\,,~~~n_{s}\simeq 0.92\,,~~~\frac{d n_s}
{d\ln k} \sim -0.002\,.
\eeq
The spectral tilt and the running agree with the current WMAP 3-years'
data within $2\sigma$~\cite{WMAP3}. The tensor modes are negligible
because of the low scale of inflation.

We emphasize that the VEV of the flat direction is related to the Yukawa
coupling that can generate the Dirac neutrino mass $\sim 0.1$~eV.
The scale of the neutrino mass appears to be just right to get the
correct amplitude in the CMB perturbations.

The inflaton has gauge couplings to the electroweak and $U(1)_{\rm
B-L}$ gauge/gaugino fields. It therefore induces a VEV-dependent
mass $\sim g \langle \phi \rangle$ for these fields ($g$ denotes a
typical gauge coupling). After the end of inflation, $\phi$ starts
oscillating around the global minimum at the origin with a frequency
$m_{\phi} \sim 10^3 H_{\rm inf}$, see eq.~(\ref{values}). When the
inflaton passes through the minimum, $\langle \phi\rangle = 0$, the
induced mass undergoes non-adiabatic time variation. This results in
non-perturbative particle production~\cite{PREHEAT}. As the inflaton
VEV is rolling back to its maximum value $\phi_0$, the mass of the
gauge/gaugino quanta increases again. Because of their large
couplings they quickly decay to the fields which are not coupled to
the inflaton, hence massless, notably the down-type (s)quarks. This
is a very efficient process as a result of which the inflaton decays
to relativistic particles within few Hubble times after the end of
inflation (for more details see~\cite{AEGJM}). A thermal bath of
MSSM particles is eventually formed with a temperature $T_{rh} \sim
10^6$ GeV (for details of thermalization in SUSY,
see~\cite{AVERDI1,AVERDI2,AVERDI3}). This is sufficiently large to
produce cold dark matter in the form of thermal
neutralinos~\cite{CDM-REV}. The temperature is also high enough for
the electroweak baryogenesis~\cite{BARYO-REV}. On the other hand,
the reheat temperature is low enough for the dangerous relics, such
as gravitinos, not to be produced~\cite{AVERDI1,AVERDI2,MAROTO}.

The inclusion of a gauged $U(1)_{\rm B-L}$ is important for
successful reheating. In the absence of it the RH (s)neutrinos would
be gauge singlet. Then the energy density in the ${\tilde N}_3$
component of the inflaton could only decay via the tiny Yukawa
coupling $h_3$, happening at a rate $\Gamma_{N} \sim (h^2_3/8 \pi)
m_{\phi} \sim 10^{-22}$ GeV. The oscillations of ${\tilde N}_3$
would dominate the universe, and their very late decay (though
before BBN, as can be seen from $\Gamma_N$) would dilute the
generated baryon asymmetry and dark matter. However, the presence of
a $U(1)_{\rm B-L}$ ensures that the energy density in the inflaton
completely decays through interactions, thus efficiently.

Some comments are in order.  In our discussion so far we have included
only one neutrino and neglected the dynamical role of other two other
generations of right handed neutrinos, which result in multiple flat
directions~\cite{JOKINEN}. Their Yukawa couplings are even smaller,
$h_{1,2} \ll 10^{-12}$.  We have checked that the presence of these
flat directions does not affect the dynamics and the temperature
anisotropy.

We have neglected so far the running of the soft supersymmetry
breaking parameters $m_\phi$ and $A$. In gravity-mediated
supersymmetry breaking models these terms run logarithmically (as
well as the Yukawa coupling $h_3$). Let us focus on the contribution
of $U(1)_{\rm B-L}$ part. The ${\rm B-L}$ charges of ${\bf N}$,
${\bf L}$ and ${\bf H}_u$ are $+1,~-1,~0$ respectively. Due to the
smallness of $h_3 \sim 10^{-12}$, the ${\rm B-L}$ gauge interactions
dominate the running. Their contribution to the one-loop
renormalization group equations of $m^2_{\phi},~A,~h_3$ closely
follows those from the $U(1)_{\rm Y}$ piece of the SM~\cite{NILLES}
\begin{eqnarray} \label{rge}
\mu {d m^2_{\phi} \over d \mu} & = & - {1 \over 12 \pi^2}~g^2_{\rm
B-L}~\vert {\tilde m} \vert^2 + ... \, , \nonumber \\
\mu {d A \over d \mu} & = & - {1 \over 8 \pi^2}~g^2_{\rm B-L}~
{\tilde m} + ... \, , \nonumber \\
\mu {d h_3 \over d \mu} & = & - {1 \over 16 \pi^2}~g^2_{\rm B-L}~h_3
+ ... \, ,
\end{eqnarray}
where $\mu$ is the scale, and $g_{\rm B-L},~{\tilde m}$ are the
gauge coupling and gaugino mass of the $U(1)_{\rm B-L}$
respectively. The "..." denotes the contributions from the SM gauge
interactions, which are similar in size. The running implies VEV
dependence of $m^2_{\phi},~A,~h_3$. This should be taken into
account when we determine the flatness condition from
eq.~(\ref{flatpot}).

Starting from the grand unified theory (GUT) scale $M_{\rm GUT}
\approx 2 \times 10^{16}$ GeV, we have
\begin{eqnarray} \label{log}
m^2_{\phi}(\phi) & = & m^2_0 ~ \Big[1 + K_1 ~ {\rm ln} \Big({\phi^2
\over
M^2_{\rm GUT}} \Big) \Big] \, , \nonumber \\
A(\phi) & = & A_0 ~ \Big[1 + K_2 ~ {\rm ln} \Big({\phi^2 \over
M^2_{\rm GUT}}
\Big) \Big] \, , \nonumber \\
h_3 (\phi) & = & h^0_3 ~ \Big[1 + K_3 ~ {\rm ln} \Big({\phi^2 \over
M^2_{\rm GUT}} \Big) \Big] \, .
\end{eqnarray}
Here $m^2_0,~A_0,~h^0_3$ are the boundary values of parameters at
$M_{\rm GUT}$ and
\begin{eqnarray} \label{k}
K_1 & \approx & - {1 \over 12 \pi^2} ~ \Big({m_{1/2} \over m_0}
\Big)^2 ~ g^2_{\rm B-L} + ... \, , \nonumber \\
K_2 & \approx & - {1 \over 8 \pi^2} ~ \Big({m_{1/2} \over A_0} \Big)
~ g^2_{\rm B-L} + ... \, , \nonumber \\
K_3 & \approx & - {1 \over 16 \pi^2} ~ g^2_{\rm B-L} + ... \, ,
\end{eqnarray}
with $m_{1/2}$ being the gaugino masses at the GUT scale. The "..."
again denotes contributions of similar size from the SM gauge
interactions.

The condition for the existence of a point $\phi_0$ such that
$V^{\prime}(\phi_0) = V^{\prime \prime} (\phi_0) = 0$, and its VEV
follow from the analysis of~\cite{AEGJM}
\begin{eqnarray} \label{cond2}
A(\phi_0) & = & 4 m_{\phi}(\phi_0) ~ \Big(1 + K_1 - {4 \over 3} K_2
+ {1 \over 2} K_3 \Big)^{1/2} \, , \nonumber \\
\phi_0 & = & {\sqrt{3} m_{\phi}(\phi_0) \over h^0_3} ~ \Big(1 + {1
\over 2} K_1 - {1 \over 4} K_3 \Big) \, .
\end{eqnarray}
Note that these expressions are the same as those given
in~(\ref{cond},\ref{phi0}) except for corrections due to running. It
can be seen from Eq.~(\ref{k}) that $K_1,~K_2,~K_3 \sim {\cal
O}(10^{-2})$. Hence the corrections are indeed very small.

However, the important point is that the condition for the flatness
of the potential involves $m_{\phi}(\phi_0),~A(\phi_0)$ rather than
the boundary values $m_0,~A_0$. One can then use renormalization
group equations to translate this into a (slightly different)
relation between $m_0$ and $A_0$. Therefore quantum corrections do
not spoil the flatness of the potential; they just slightly modify
the flatness condition. In consequence, the rest of our analyses
remains unchanged: the number of e-foldings~(\ref{totalefold}), the
amplitude of perturbations~(\ref{amp}), etc. One should just note
that the values of parameters at $\phi_0$ are used in respective
expressions~\footnote{Note that inflation takes place in an interval
$\Delta \phi \sim 10^{-8} \phi_0$ around $\phi_0$, see
Eq.~(\ref{end}). Therefore logarithmic running of parameters in this
interval is totally negligible. The only relevant running is that
from the GUT scale down to $\phi_0 \sim 10^{14}$ GeV.}.

Finally the issue of initial condition could be addressed by
triggering early bouts of inflation as has been often argued, see
e.g. Refs.~\cite{BURGESS,AFM}, with a last phase driven by the
A-term inflaton. Further note that near the saddle point of an
A-term inflation, there is a self-reproduction eternal inflation
regime, as we discussed above (see eq.~(\ref{drift}) and the
discussions after eq.~(\ref{values})). The eternal inflation regime
along with a prior phase of a false vacuum inflation ameliorates the
initial condition problem considerably for a cosmologically flat
direction as we have discussed in Ref.~\cite{AFM}.

To summarize, in the MSSM with right-handed neutrinos the flat
direction parameterized by the VEV of the right-handed sneutrino, the
slepton, and the Higgs field can serve as the inflaton.  To produce
the correct density perturbations, the neutrino Yukawa coupling should
be small, and the scale of the corresponding Dirac masses turns out to
be in general agreement with the known neutrino masses.  The salient
feature of the present model is that the inflaton is not an additional
{\em ad hoc} ingredient; it is firmly rooted in particle physics, and
its properties can be inferred from the upcoming experiments,
including the LHC.

% \vspace{15pt}

%%%%%%%%%%%%%%%%%%%%%%%%%%%%%%%%%%%%%%%%%%%%%%%%%%%%%%%%%%%%%%%%%%%%%%%%%%%%%%%
% {\it Acknowledgments-}
The work of R.A. is supported by the National
Sciences and Engineering Research Council of Canada (NSERC).  The work
of A.K. was supported in part by the U.S. Department of Energy grant
DE-FG03-91ER40662 and by the NASA ATP grants NAG~5-10842 and
NAG~5-13399.

%%%%%%%%%%%%%%%%%%%%%%%%%%%%%%%%%%%%%%%%%%%%%%%%%%%%%%%%%%%%%%%%%%%%%%%%%%%%%%%


\begin{thebibliography}{99}

\bibitem{AEGM}
R.~Allahverdi, K.~Enqvist, J.~Garcia-Bellido and A.~Mazumdar,
  %``Gauge invariant MSSM inflaton,''
  Phys.\ Rev.\ Lett.\  {\bf 97}, 191304 (2006)
  [arXiv:hep-ph/0605035].
  %%CITATION = PRLTA,97,191304;%%


\bibitem{LHC}
See for instance: http://lhc.web.cern.ch/lhc/

\bibitem{WMAP3}
D.N. Spergel, et.al., astro-ph/0603449.

\bibitem{FEW}
G.~Lazarides and Q.~Shafi,
  %``A Predictive inflationary scenario without the gauge singlet,''
  Phys.\ Lett.\ B {\bf 308}, 17 (1993).
  %%CITATION = HEP-PH 9304247;%%
S.~Kasuya, T.~Moroi and F.~Takahashi,
  %``Can MSSM particle be the inflaton?,''
  Phys.\ Lett.\ B {\bf 593}, 33 (2004).
%\bibitem{ASKO}
 R.~Brandenberger, P.~M.~Ho and H.~C.~Kao,
  %``Large N cosmology,''
  JCAP {\bf 0411}, 011 (2004).
  %%CITATION = HEP-TH 0312288;%%
A.~Jokinen and A.~Mazumdar,
  %``Inflation in large N limit of supersymmetric gauge theories,''
  Phys.\ Lett.\ B {\bf 597}, 222 (2004).
  %%CITATION = HEP-TH 0406074;%%

%\bibitem{ASSIST}
% A.~R.~Liddle, A.~Mazumdar and F.~E.~Schunck,
  %``Assisted inflation,''
%  Phys.\ Rev.\ D {\bf 58}, 061301 (1998).
  %%CITATION = ASTRO-PH 9804177;%%
% E.~J.~Copeland, A.~Mazumdar and N.~J.~Nunes,
  %``Generalized assisted inflation,''
%  Phys.\ Rev.\ D {\bf 60}, 083506 (1999).
  %%CITATION = ASTRO-PH 9904309;%%
%A.~Mazumdar, S.~Panda and A.~Perez-Lorenzana,
%``Assisted inflation via tachyon condensation,''
%Nucl.\ Phys.\ B {\bf 614}, 101 (2001).
%%CITATION = HEP-PH 0107058;%%


\bibitem{seesaw} P. Minkowski, Phys. lett. {\bf B67 }, 421
(1977); M.~Gell-Mann, P.~Ramond, and R.~Slansky, \emph{Supergravity}
(P.~van Nieuwenhuizen et al. eds.), North Holland, Amsterdam, 1980,
p.~315; T.~Yanagida, in \emph{Proceedings of the Workshop on the
Unified Theory and the Baryon Number in the Universe} (O.~Sawada
and A.~Sugamoto, eds.), KEK, Tsukuba, Japan, 1979, p.~95; S.~L.
Glashow, \emph{The future of elementary particle physics}, in
   \emph{Proceedings of the 1979 Carg{\`e}se Summer Institute on
Quarks and Leptons} (M.~L{\'e}vy et al. eds.), Plenum Press, New York,
1980, pp.~687; R.~N. Mohapatra and G.~Senjanovi{\'c}, Phys. Rev. Lett.
\textbf{44}, 912 (1980).
  %%CITATION = PRLTA,44,912;%%


\bibitem{NEUT-REV}
 For review, see, e.g., R.~N.~Mohapatra {\it et al.},
%   ``Theory of neutrinos: A white paper,''
  arXiv:hep-ph/0510213.
  %%CITATION = HEP-PH 0510213;%%




 \bibitem{deGouvea:2005er}
  A.~de Gouvea,
  %``See-saw energy scale and the LSND anomaly,''
  Phys.\ Rev.\ D {\bf 72}, 033005 (2005).
%   [arXiv:hep-ph/0501039].
  %%CITATION = HEP-PH 0501039;%%

\bibitem{nuMSM}
 T.~Asaka, S.~Blanchet and M.~Shaposhnikov,
  %``The nuMSM, dark matter and neutrino masses,''
  Phys.\ Lett.\ B {\bf 631}, 151 (2005);
  %[arXiv:hep-ph/0503065].
  %%CITATION = HEP-PH 0503065;%%



  \bibitem{dark_matter}
S.~Dodelson and L.~M.~Widrow,
%``Sterile-neutrinos as dark matter,''
Phys.\ Rev.\ Lett.\  {\bf 72}, 17 (1994).
%[arXiv:hep-ph/9303287].
%%CITATION = HEP-PH 9303287;%%
%
% \bibitem{Fuller}
K.~Abazajian, G.~M.~Fuller and M.~Patel,
%``Sterile neutrino hot, warm, and cold dark matter,''
Phys.\ Rev.\ D {\bf 64}, 023501 (2001);
%[arXiv:astro-ph/0101524].
%%CITATION = ASTRO-PH 0101524;%%
%
% \bibitem{dolgov_hansen}
A.~D.~Dolgov and S.~H.~Hansen,
%``Massive sterile neutrinos as warm dark matter,''
Astropart.\ Phys.\  {\bf 16}, 339 (2002);
%[arXiv:hep-ph/0009083].
%%CITATION = HEP-PH 0009083;%%
  M.~Shaposhnikov,
  %``A possible symmetry of the nuMSM,''
  arXiv:hep-ph/0605047;
  %%CITATION = HEP-PH 0605047;%%
  A.~Kusenko,
  %``Sterile neutrinos, dark matter, and the pulsar velocities in models  with
% a Higgs singlet,''
 Phys.\ Rev.\ Lett.\  {\bf 97}, 241301 (2006)
%   [arXiv:hep-ph/0609081].
  %%CITATION = PRLTA,97,241301;%%

\bibitem{pulsars}
  A.~Kusenko and G.~Segr\`e,
%``Neutral current induced neutrino oscillations in a supernova,''
Phys.\ Lett.\ B {\bf 396}, 197 (1997);
%[arXiv:hep-ph/9701311].
%%CITATION = HEP-PH 9701311;%%
A.~Kusenko and G.~Segre,
  %``Pulsar kicks from neutrino oscillations,''
  Phys.\ Rev.\ D {\bf 59}, 061302 (1999);
%   [arXiv:astro-ph/9811144].
  %%CITATION = ASTRO-PH 9811144;%%
%
% \bibitem{fkmp}
G.~M.~Fuller, A.~Kusenko, I.~Mocioiu, and S.~Pascoli,
%``Pulsar kicks from a dark-matter sterile neutrino,''
Phys.\ Rev.\ D {\bf 68}, 103002 (2003).
%%CITATION = ASTRO-PH 0307267;%%
For review, see, {\em e.g.},
  A.~Kusenko,
  %``Pulsar kicks from neutrino oscillations,''
  Int.\ J.\ Mod.\ Phys.\ D {\bf 13}, 2065 (2004);
  %[arXiv:astro-ph/0409521].
  %%CITATION = ASTRO-PH 0409521;%%
M.~Barkovich, J.~C.~D'Olivo and R.~Montemayor,
%``Active-sterile neutrino oscillations and pulsar kicks,''
Phys.\ Rev.\ D {\bf 70}, 043005 (2004).
%[arXiv:hep-ph/0402259].
%%CITATION = HEP-PH 0402259;%%



\bibitem{reion}
  P.~L.~Biermann and A.~Kusenko,
  %``Relic keV sterile neutrinos and reionization,''
 Phys.\ Rev.\ Lett.\  {\bf 96}, 091301 (2006);
  %%CITATION = ASTRO-PH 0601004;%%
%
% \bibitem{reion1}
  J.~Stasielak, P.~L.~Biermann and A.~Kusenko,
  %``Thermal evolution of the primordial clouds in warm dark matter models with
  %keV sterile neutrinos,''
  arXiv:astro-ph/0606435.
  %%CITATION = ASTRO-PH 0606435;%%



  \bibitem{baryogenesis}
  E.~K.~Akhmedov, V.~A.~Rubakov and A.~Y.~Smirnov,
  %``Baryogenesis via neutrino oscillations,''
  Phys.\ Rev.\ Lett.\  {\bf 81}, 1359 (1998);
%  [arXiv:hep-ph/9803255].
  %%CITATION = HEP-PH 9803255;%%
%\bibitem{Asaka:2005pn_baryons}
  T.~Asaka and M.~Shaposhnikov,
  %``The nuMSM, dark matter and baryon asymmetry of the universe,''
  Phys.\ Lett.\ B {\bf 620}, 17 (2005).
%  [arXiv:hep-ph/0505013].
  %%CITATION = HEP-PH 0505013;%%




\bibitem{DRT}
M.~Dine, L.~Randall and S.~Thomas, Phys. Rev. Lett. {\bf 75}, 398 (1995).
M. Dine, L. Randall and S. Thomas, Nucl. Phys. B {\bf 458}, 291 (1996).
T. Gherghetta, C. Kolda and S. P. Martin, Nucl. Phys. B {\bf 468},
37 (1996).
%%CITATION = HEP-PH 9510370;%%
%



\bibitem{MSSM-REV}
For reviews, see
 K.~Enqvist and A.~Mazumdar,
  %``Cosmological consequences of MSSM flat directions,''
  Phys.\ Rept.\  {\bf 380}, 99 (2003);
  %%CITATION = HEP-PH 0209244;%%
 M. Dine and A. Kusenko, Rev. Mod. Phys. {\bf 76}, 1 (2004).
%%CITATION = HEP-PH 0303065;%%


\bibitem{curvaton}
R.~Allahverdi, K.~Enqvist, A.~Jokinen and A.~Mazumdar,
  %``Identifying the curvaton within MSSM,''
  JCAP {\bf 0610}, 007 (2006)
  [arXiv:hep-ph/0603255].
  %%CITATION = JCAPA,0610,007;%%


\bibitem{DHL}
D.~H.~Lyth,
  arXiv:hep-ph/0605283.
  %%CITATION = HEP-PH 0605283;%%
 R.~Allahverdi and A.~Mazumdar,
  %``Spectral tilt in A-term inflation,''
  arXiv:hep-ph/0610069.
  %%CITATION = HEP-PH/0610069;%%

\bibitem{GUTH}
A.~H.~Guth,
  %``The Inflationary Universe: A Possible Solution To The Horizon And Flatness
  %Problems,''
 Phys.\ Rev.\ D {\bf 23}, 347 (1981).
  %%CITATION = PHRVA,D23,347;%%


\bibitem{AEGJM}
 R.~Allahverdi, K.~Enqvist, J.~Garcia-Bellido, A.~Jokinen and A.~Mazumdar,
  %``MSSM flat direction inflation: Slow roll, stability, fine tunning and
  %reheating,'' JCAP {\bf 06}, 019 (2007)
  arXiv:hep-ph/0610134.
  %%CITATION = HEP-PH/0610134;%%


\bibitem{LINDE}
  A.~D.~Linde,
%   ``Particle Physics and Inflationary Cosmology,''
  Contemp.\ Concepts Phys.\  {\bf 5}, 1 (2005).
  [arXiv:hep-th/0503203].
  %%CITATION = HEP-TH 0503203;%%


\bibitem{LEACH}
A.~R.~Liddle and S.~M.~Leach,
  %``How long before the end of inflation were observable perturbations
  %produced?,''
  Phys.\ Rev.\ D {\bf 68}, 103503 (2003).
  %%CITATION = ASTRO-PH 0305263;%%

\bibitem{PREHEAT}
L. Kofman, A. S. Linde and A. A. Starobinsky, Phys. Rev. Lett. {\bf
73}, 3195 (1994); L. Kofman, A. D. Linde and A. A. Starobinsky,
Phys. Rev. D {\bf 56}, 3258 (1997).


\bibitem{AVERDI1}
 R.~Allahverdi and A.~Mazumdar,
  %``Quasi-thermal universe: From cosmology to colliders,''
  arXiv:hep-ph/0505050.
  %%CITATION = HEP-PH 0505050;%%


\bibitem{AVERDI2}
R.~Allahverdi and A.~Mazumdar,
  %``Supersymmetric thermalization and quasi-thermal universe: Consequences  for
  %gravitinos and leptogenesis,''
  JCAP {\bf 0610}, 008 (2006)
  [arXiv:hep-ph/0512227].
  %%CITATION = JCAPA,0610,008;%%


\bibitem{AVERDI3}
 R.~Allahverdi and A.~Mazumdar,
  %``Towards a successful reheating within supersymmetry,''
  arXiv:hep-ph/0603244.
  %%CITATION = HEP-PH 0603244;%%


\bibitem{CDM-REV}
 G.~Bertone, D.~Hooper and J.~Silk,
  %``Particle dark matter: Evidence, candidates and constraints,''
  Phys.\ Rept.\  {\bf 405}, 279 (2005).
  %%CITATION = HEP-PH 0404175;%%


\bibitem{BARYO-REV}
V.~A.~Rubakov and M.~E.~Shaposhnikov,
  %``Electroweak baryon number non-conservation in the early universe and in
  %high-energy collisions,''
  Usp.\ Fiz.\ Nauk {\bf 166}, 493 (1996)
  [Phys.\ Usp.\  {\bf 39}, 461 (1996)];
  %%CITATION = HEP-PH 9603208;%%
J.~Garcia-Bellido, D.~Y.~Grigoriev, A.~Kusenko and M.~E.~Shaposhnikov,
  %``Non-equilibrium electroweak baryogenesis from preheating after  inflation,''
  Phys.\ Rev.\ D {\bf 60} (1999) 123504.
  %%CITATION = HEP-PH 9902449;%%

\bibitem{MAROTO}
 A.~L.~Maroto and A.~Mazumdar,
%   ``Production of spin 3/2 particles from vacuum fluctuations,''
  Phys.\ Rev.\ Lett.\  {\bf 84}, 1655 (2000)
%  [arXiv:hep-ph/9904206].
  %%CITATION = HEP-PH 9904206;%%
 G.~F.~Giudice, A.~Riotto and I.~Tkachev,
%   ``Thermal and non-thermal production of gravitinos in the early universe,''
  JHEP {\bf 9911}, 036 (1999)
%  [arXiv:hep-ph/9911302].
  %%CITATION = HEP-PH 9911302;%%


\bibitem{JOKINEN}
 K.~Enqvist, A.~Jokinen and A.~Mazumdar,
  %``Dynamics of MSSM flat directions consisting of multiple scalar fields,''
  JCAP {\bf 0401}, 008 (2004).
  %%CITATION = HEP-PH 0311336;%%

\bibitem{NILLES}

H. P. Nilles, Phys. Rept. {\bf 110}, 1 (1984).

\bibitem{BURGESS}
C.~P.~Burgess, R.~Easther, A.~Mazumdar, D.~F.~Mota and T.~Multamaki,
  %``Multiple inflation, cosmic string networks and the string landscape,''
  JHEP {\bf 0505}, 067 (2005).
  %%CITATION = HEP-TH 0501125;%%

\bibitem{AFM} R. Allahverdi, A. R. Frey ad A. Mazumdar,
arXiv:hep-th/0701233 (To be published in Phys. Rev. D).





\end{thebibliography}
\end{document}